\documentclass{article}
\usepackage{acra}
\usepackage{booktabs}
\usepackage{graphicx}
\usepackage{url}

\title{Exploring Mixed Reality for Engagement in Robotics: A Comparison of Robot Control Methods}

\author{Samantha Edvell, Dinh Tung Le, and Gavin Paul \\
UTS Robotics Institute, University of Technology Sydney, Australia \\
\texttt{Samantha.E.Edvell@student.uts.edu.au}}

\begin{document}
\maketitle

\begin{abstract}
This paper presents a Mixed Reality (MR) robotics experience that provides an accessible introduction to interacting with a physical robot. The application combines a TurtleBot with a virtual animal-tending activity and three control methods: \textit{Joystick}, \textit{Waypoint}, and \textit{Path-Drawing}. A mixed-methods user study with 10 participants investigated changes in robotics interest and confidence following the experience and how the control methods influenced engagement, workload, and preference. Confidence increased following the experience, with the exploratory analysis indicating a significant pre--post difference, while the increase in interest was not statistically significant. The control methods produced clear differences in perceived workload and ease of use, with \textit{Path-Drawing} generally more demanding and difficult to use. \textit{Joystick} was preferred by eight of ten participants and was associated with better perceived performance and lower frustration. However, no statistically significant differences were detected in the other engagement dimensions, suggesting that the most preferred method was not necessarily more engaging across all aspects of the experience. These findings highlight the importance of considering both usability and engagement when designing how users interact with robots in introductory MR experiences.
\end{abstract}

\section{Introduction}
\label{sec:introduction}

\begin{figure}[ht]
    \centering
    \includegraphics[width=\columnwidth]{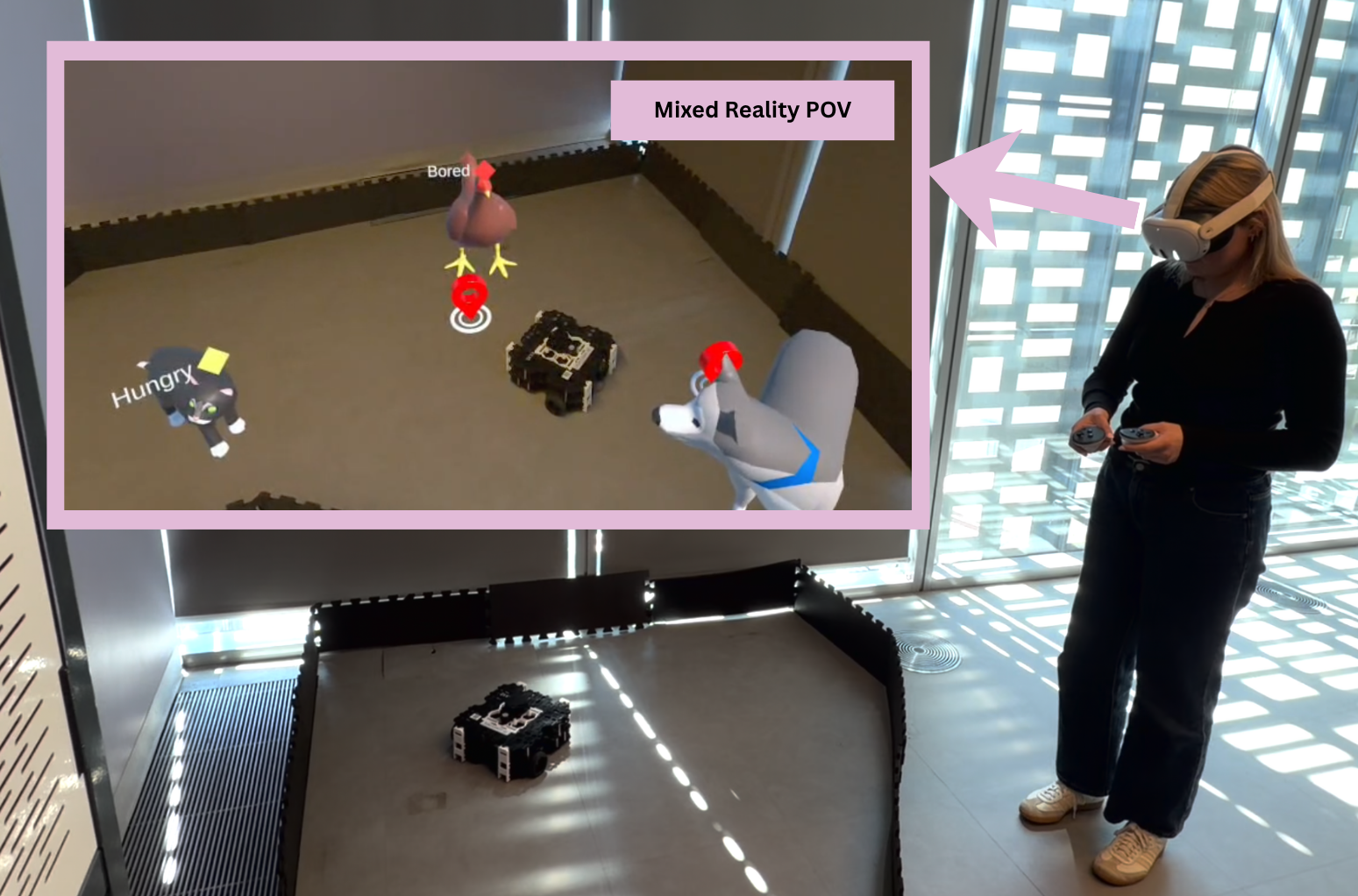}
    \caption{Participant controlling the TurtleBot through the MR application, with the mixed-reality view overlaid. Published with the participant's explicit consent.}
    \label{fig:open_photo}
\end{figure}

As robots become increasingly integrated into industry, education, and everyday environments, opportunities to understand and engage with robotics must be accessible to a broad range of people. However, women remain underrepresented in robotics~\cite{cimpian_understanding_2020,fields_underrepresentation_2024}. Providing girls with opportunities to interact with robotics may encourage future participation in the field. Previous robotics programs for girls have reported positive outcomes, including increased interest, confidence, STEM identity, and aspirations towards further STEM learning~\cite{mich_young_2019,yabas_empowering_2022,maigaard_kjar_pedersen_using_2021}. Furthermore, in one primary-school robotics study, girls' initially lower self-efficacy increased following the activity~\cite{tatnall_gender_2017}. This highlights the importance of providing opportunities for girls to gain experience and build confidence in STEM activities.

 Creating these opportunities requires consideration of not only what participants are exposed to, but how they experience that interaction. Mixed Reality (MR) provides one approach by allowing virtual content to be integrated with interaction with a physical robot. Extended Reality (XR) has been associated with curiosity, enjoyment, self-efficacy, and intrinsic motivation in educational settings~\cite{crogman_virtual_2025,mystakidis_immersive_2023}. Existing applications such as RoboStage combined MR with embodied robot characters and reported positive effects on perceived authenticity and learning motivation~\cite{Chang2010RoboStage}. MR therefore provides a promising medium for creating introductory robotics experiences for users with limited previous exposure.
 
However, using MR does not itself ensure that an experience will be engaging. Previous research indicates that the design and framing of robotics activities can influence girls' engagement~\cite{voutyrakou_using_2025}. Accordingly, this work develops an MR robotics experience motivated by the need to provide girls with limited prior robotics experience an accessible and engaging introduction to the field, while also investigating how the method of robot control shapes that experience.
 
 This work presents an MR application in which users interact with a physical TurtleBot3 Waffle Pi through an animal-tending activity. Three control methods---joystick control (\textit{Joystick}), waypoint placement (\textit{Waypoint}), and path drawing (\textit{Path-Drawing})---are incorporated into the experience and evaluated through a user study that considers robotics interest and confidence, engagement, perceived workload, and preference.

The intended contributions of this paper are:
\begin{itemize}
\item the design and implementation of an accessible MR robotics experience, informed by research on engaging girls in STEM;
\item a comparison of three levels of robot control abstraction within the same MR experience; and
\item findings on how the experience relates to robotics interest and confidence, and how the control abstraction methods influence engagement, perceived workload, and preference.
\end{itemize}

\section{Related Work}
\label{sec:related_work}

\subsection{Designing Engaging Educational Robotics Activities for Girls}

The design and construction of educational robotics activities can influence their effectiveness~\cite{voutyrakou_using_2025}. Previous research has identified several factors that may contribute to girls' engagement. Reviews recommend collaborative rather than competitive activities to support girls' participation~\cite{moro_how_2020}, and collaborative strategies have also been observed in group robotics activities~\cite{merdan_participants_2020}. These findings support the consideration of collaborative and non-competitive approaches when designing robotics activities intended to engage girls. Studies involving girls have similarly identified direct human--robot interaction as beneficial and preferred~\cite {maigaard_kjar_pedersen_using_2021}. Engagement may also be increased when STEM outreach activities emphasise meaningful real-world or societal applications~\cite{maigaard_kjar_pedersen_using_2021,baizan_activities_2021}. Together, these findings suggest that providing active interaction within a collaborative, meaningful context may be valuable when designing robotics experiences intended to engage girls.

These findings should not be interpreted as universal gender-based preferences, as STEM experiences and preferences vary between individuals~\cite{jakobi_framework_2026,dirsuweit_barriers_2023}. Instead, they provide design considerations that informed the use of active robot control, a shared non-competitive objective, and a meaningful narrative context in the present work.

\subsection{Design Considerations for Mixed Reality Robotics Activities}

The flexibility of MR creates a broad range of possibilities for how users interact with robots, making interaction design an important consideration. Walker et al. describe a broad design space for Virtual, Augmented, and Mixed Reality human--robot interaction, encompassing elements relating to the robot, task, environment, and other agents~\cite{walker_virtual_2023}. This flexibility creates opportunities to tailor MR robotics experiences to particular users and objectives.

However, the increased visual information in an immersive environment can increase cognitive load and make relevant information more difficult to process~\cite{makransky_cognitive_2021}. Similarly, improvements observed in immersive robotics applications have not necessarily occurred across all measured learning outcomes~\cite{OuYang2023ARBot}. At the same time, research suggests that deliberate application design can reduce these drawbacks, while immersive elements can support curiosity when incorporated meaningfully into the experience~\cite{makransky_cognitive_2021,crogman_virtual_2025}. Robot control is one such design consideration, with methods varying in their level of direct interaction and autonomy. These findings highlight the importance of considering how specific design choices influence different dimensions of the user experience, rather than assuming that MR use itself will produce uniformly positive outcomes.

\section{System Design}
\label{sec:design}

The developed system combines MR with a physical mobile robot to provide an introductory robotics experience in which users interact directly with both virtual and physical elements.

\subsection{Design Rationale}
\label{sec:design_rationale}

The MR application was designed as an accessible and engaging introductory robotics experience, informed by the design considerations identified in Section~\ref{sec:related_work}. It combines active interaction with a physical robot, a shared non-competitive objective, and an animal-themed narrative. Three control methods with different levels of control abstraction were incorporated to explore how robot control influences the user experience, as described in Section~\ref{sec:control_methods}.

\subsection{MR Environment and Task}
\label{sec:mr_task}

The MR application uses colour passthrough, allowing participants to observe the physical environment including the TurtleBot, while virtual game elements are placed within the same space. The virtual elements allowed interaction objectives to be created and modified in software without requiring additional physical objects or changes to the robot's environment.

\begin{figure}[ht]
    \centering
    \includegraphics[width=\columnwidth]{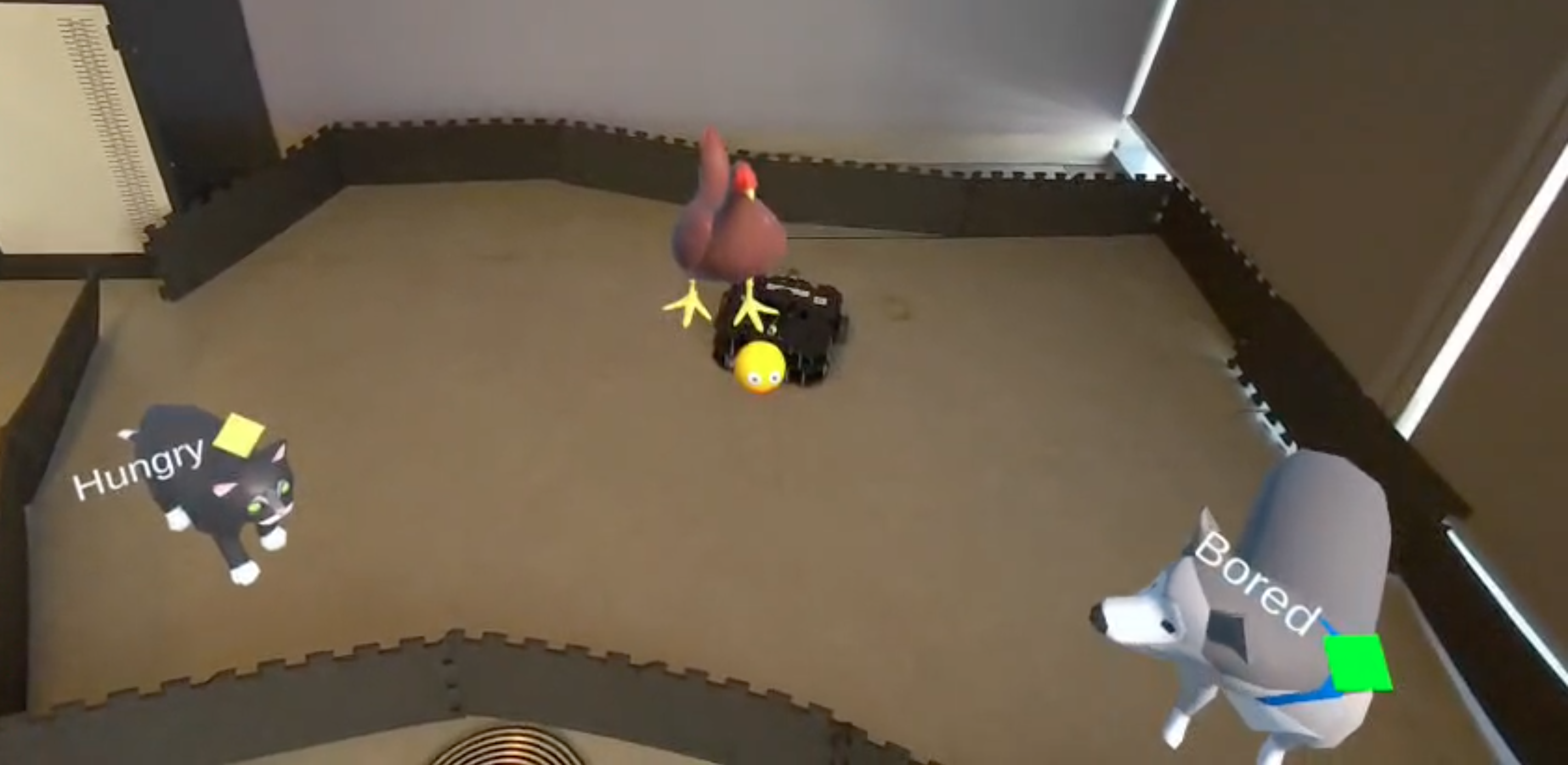}
    \caption{MR view showing virtual animals displaying their current needs and a need being resolved when the robot approaches.}
    \label{fig:animal_needs}
\end{figure}

The primary task requires participants to navigate the TurtleBot through the environment to interact with virtual animals \footnote{Animal models from the Unity Asset Store: \url{https://assetstore.unity.com/packages/3d/characters/animals/animals-free-animated-low-poly-3d-models-260727}} positioned throughout the MR space. Each animal displays a need above its head---hungry, thirsty, or bored---which changes in colour as the need becomes more urgent over time. When the robot enters an animal's interaction bounds, the corresponding need is automatically resolved by displaying food, water, or a toy \footnote{Virtual food, water, and toy assets were obtained from the Unity Asset Store: \url{https://assetstore.unity.com/packages/3d/props/food/free-hyper-casual-street-food-pack-283755}, \url{https://assetstore.unity.com/packages/3d/props/small-kit-3d-stylized-petshop-asset-291990}, and \url{https://assetstore.unity.com/packages/3d/props/low-poly-objects-pack-241890}, respectively.} as shown in Figure~\ref{fig:animal_needs}. 

After a randomly timed interval, a new need appears, requiring participants to continue navigating between animals throughout the task. The activity did not include a score or leaderboard, keeping the objective non-competitive.

\subsection{Robot Control Methods}
\label{sec:control_methods}

\begin{figure*}[t]
    \centering
    \begin{tabular}{ccc}
    \includegraphics[height=5cm]{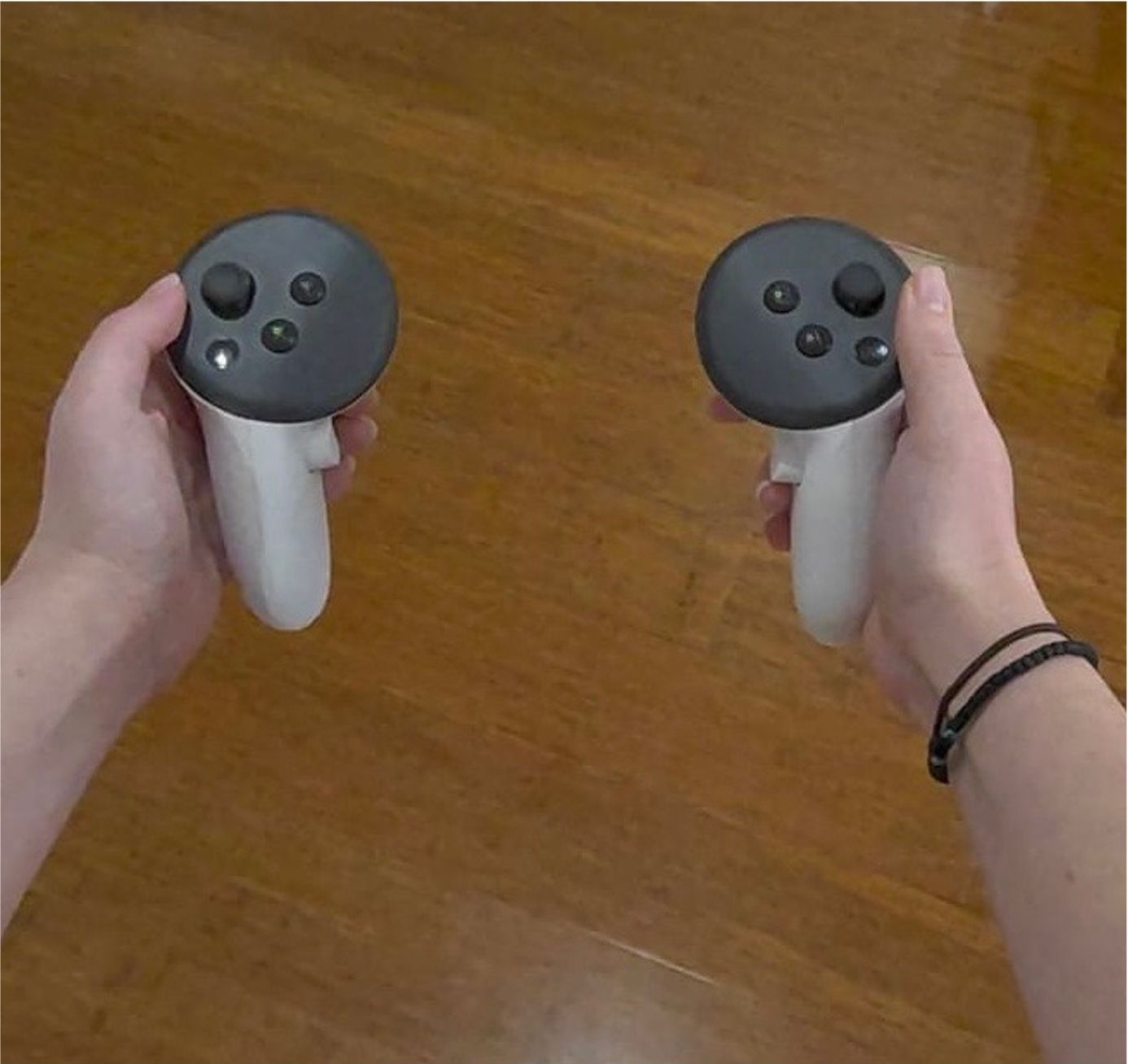} &
    \includegraphics[height=5cm]{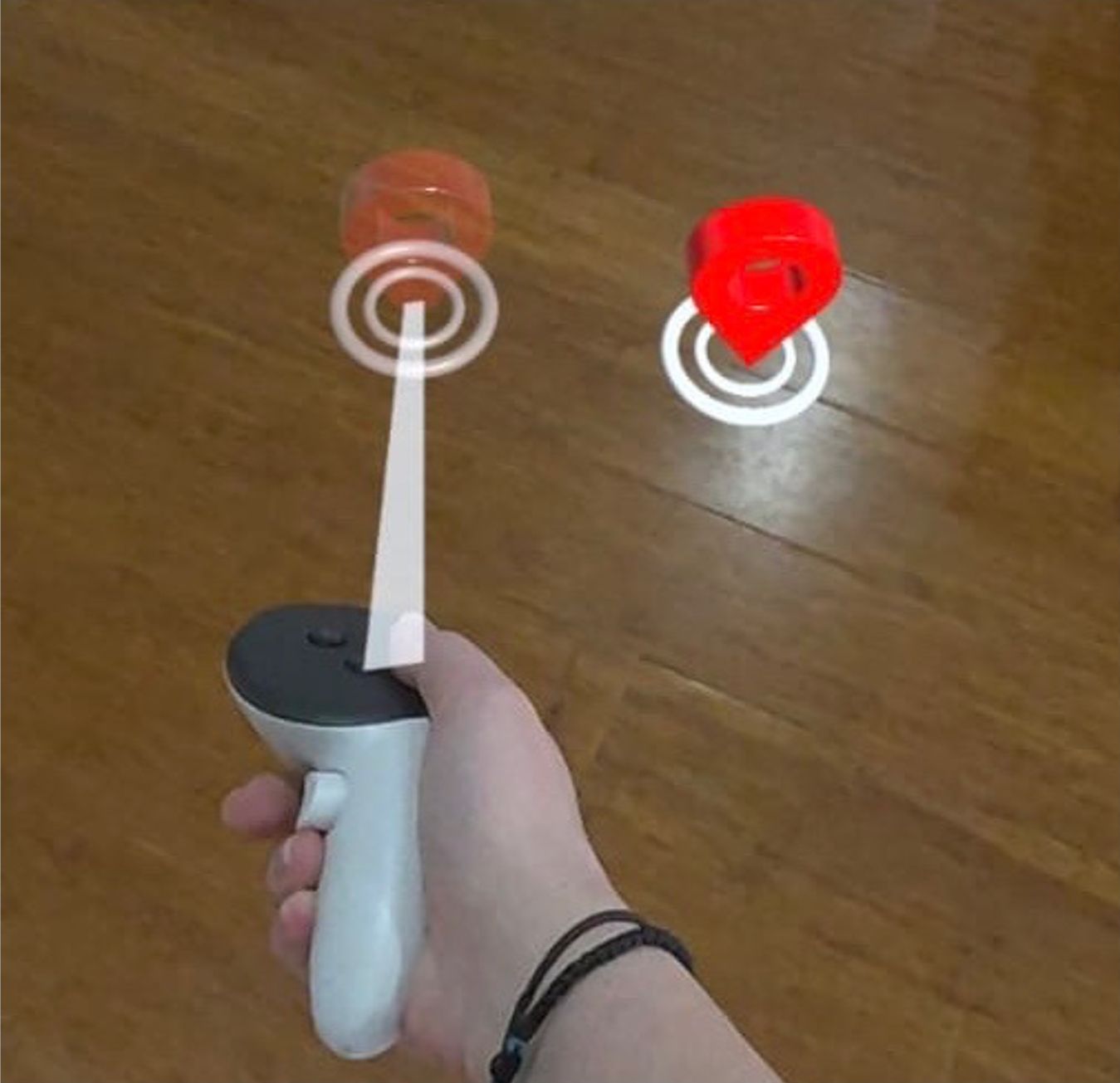} &
    \includegraphics[height=5cm]{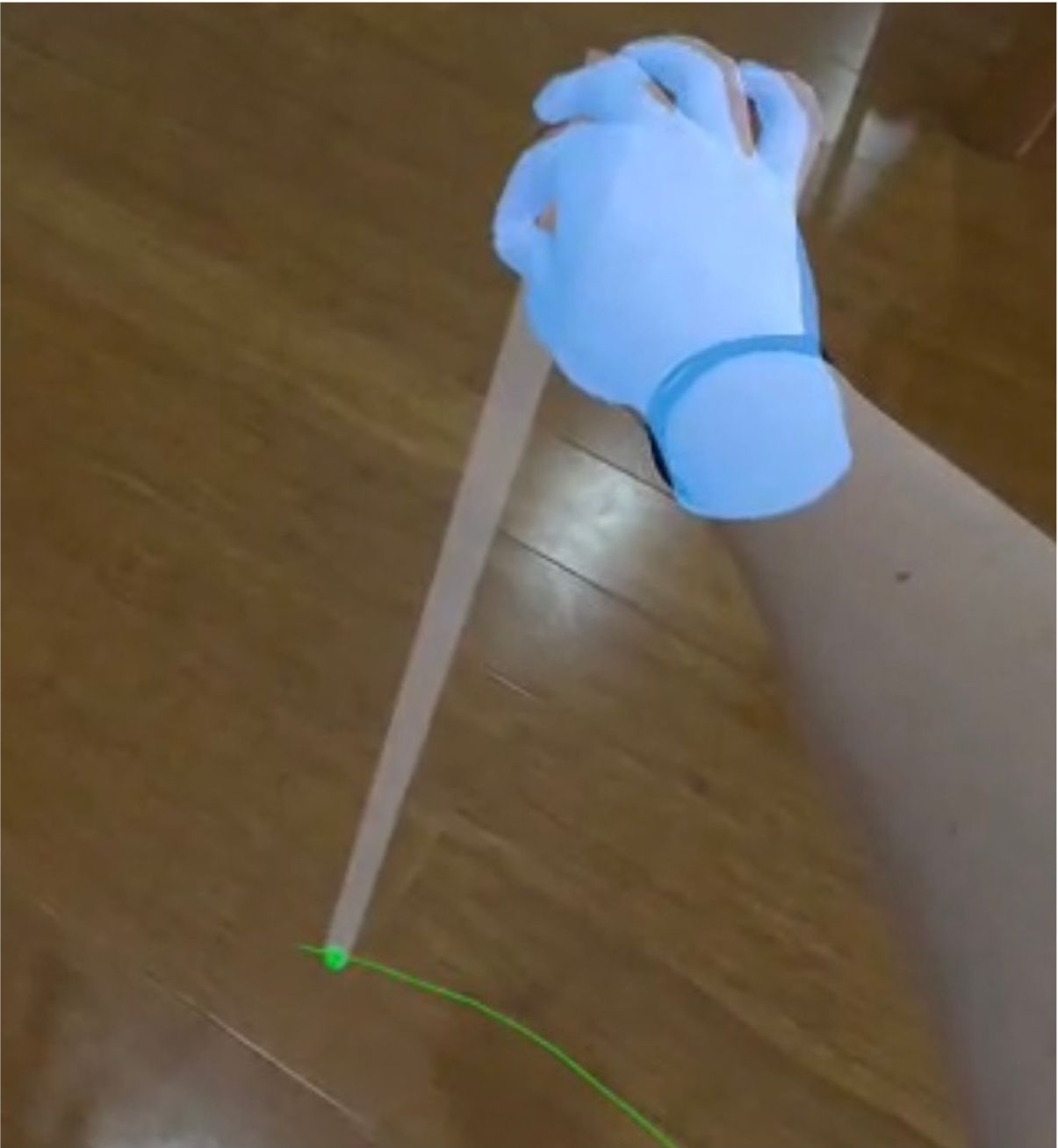} \\
    (a) \textit{Joystick} & (b) \textit{Waypoint} & (c) \textit{Path-Drawing}
    \end{tabular}
    \caption{MR interfaces for the three robot control methods.}
    \label{fig:control_methods}
\end{figure*}

To complete the task described above, participants could control the TurtleBot using three interaction methods: \textit{Joystick}, \textit{Waypoint}, and \textit{Path-Drawing} (see Figure~\ref{fig:control_methods}). These methods offer different ways to communicate motion intentions to the robot and vary in the level of abstraction between participant input and physical robot motion.

\subsubsection{Joystick Control}

The \textit{Joystick} control method provides continuous direct control of the TurtleBot using the Meta Quest controllers. Participants use the right joystick to control forward and backward motion and the left joystick to turn the robot. Participants therefore continuously control the robot's direction and movement while navigating between animals.

\subsubsection{Waypoint Control}

The \textit{Waypoint} control method allows participants to select a destination within the MR environment using the Meta Quest controller. Participants point the controller toward the desired location, and a virtual marker previews the selected position. Pressing the designated controller button places the waypoint, after which the TurtleBot autonomously navigates to the selected location. This allows participants to specify where the robot should travel without directly controlling its movement along the route.

\subsubsection{Path-Drawing Control}
\textit{Path-Drawing} control allows participants to specify a desired route using hand tracking. A virtual pointer previews the starting position. Participants activate path drawing by pinching their index finger and thumb together, then move their hand through the MR environment to create the desired path. Once completed, the TurtleBot follows the specified route autonomously.

\subsection{System Architecture}
\label{sec:architecture}

The system consists of the Meta Quest~3 MR interface, a Unity~6 application, and the ROS~2-based TurtleBot3 Waffle Pi. Unity manages the virtual environment and participant interactions, while ROS~2 manages robot localisation, control, and autonomous navigation. Communication between Unity and ROS~2 is provided through the ROS--Unity TCP Connector and Endpoint packages.\footnote{Unity Robotics Hub: \url{https://github.com/Unity-Technologies/Unity-Robotics-Hub}}

To align the Unity and ROS~2 coordinate systems, an AprilTag placed on the TurtleBot is observed by the Meta Quest~3 and processed using the \texttt{jp.keijiro.apriltag} package.\footnote{AprilTag package for Unity: \url{https://github.com/keijiro/jp.keijiro.apriltag}} The robot's corresponding pose within the ROS~2 map is obtained through the ROS~2 TF tree and used as a common reference to determine the transformation between the two coordinate systems. Once calibrated, the AprilTag is removed and the stored transformation is used to convert spatial information between Unity and ROS~2.

Unity sends participant inputs to ROS~2 according to the selected control method. Joystick input directly controls the robot's velocity, while waypoint and path-drawing inputs specify a destination goal or path that is executed autonomously using Nav2.\footnote{Nav2: \url{https://github.com/ros-navigation/navigation2}} When an animal is spawned, its position and interaction radius are recorded in a ROS~2 node, where they are compared with the robot's position during operation. When the robot enters an animal's interaction zone, ROS~2 notifies Unity to trigger the corresponding virtual interaction.

\section{User Study}
\label{sec:study}

A mixed-methods user study was conducted to evaluate the developed MR robotics experience. Quantitative measures assessed changes in self-reported robotics interest and confidence and compared engagement and workload across control methods, while qualitative responses provided further insight into participants' experiences and preferences.

The study addressed the following research questions:

\textbf{RQ1:} Among participants with varying levels of prior robotics experience, how do self-reported interest and confidence in learning more about robotics change from before to after completing an MR robotics experience?

It was hypothesised that participants would report increased interest and confidence in learning more about robotics after completing the experience, particularly among participants without prior robotics experience.

\textbf{RQ2:} Among participants with varying levels of prior robotics experience, how do different methods of control abstraction differ in perceived engagement, workload, and preference?

It was hypothesised that \textit{Joystick} would be preferred due to the directness of the interaction. It was additionally hypothesised that participants with prior robotics experience may show a greater preference for \textit{Path-Drawing} control due to the novelty of the interaction.

Engagement and workload were considered together to provide a broader understanding of how different aspects of the interaction shape the user experience.

\subsection{Participants}
\label{sec:participants}

A total of 10 volunteers (7 female, 3 male), aged 20--22 years ($M=21.6$, $SD=0.70$), participated in the study. Five participants had prior robotics experience. Prior headset experience varied, with one participant reporting no previous use, six reporting 1--2 uses, two reporting 3--10 uses, and one reporting more than 10 uses.

Ethics approval was granted by the University of Technology Sydney, Australia (ETH21-5929). All participants provided informed consent before taking part.

\subsection{Study Design and Procedure}
\label{sec:procedure}
A within-subjects design was used to compare the three control methods, with every participant experiencing \textit{Joystick}, \textit{Waypoint}, and \textit{Path-Drawing} control.

Before interacting with the system, participants completed a pre-study questionnaire recording their prior experience with robotics and immersive technologies and their initial attitudes towards robotics. Participants rated their agreement with five statements (see Table~\ref{tab:interest_confidence_statements}) on a five-point Likert scale.
\begin{table}[t]
\centering
\caption{Statements used to assess participants' interest and confidence in robotics.}
\label{tab:interest_confidence_statements}
\small
\begin{tabular}{p{0.4cm} p{1.4cm} p{5.5cm}}
\hline
\textbf{ID} & \textbf{Category} & \textbf{Statement} \\
\hline
S1 & Interest & I am interested in learning more about robotics. \\
S2 & Interest & I would like to learn how to control a robot. \\
S3 & Confidence & I feel capable of learning to control a robot. \\
S4 & Interest & I would like to learn more about how robots work. \\
S5 & Confidence & I would feel confident trying a new robotics activity. \\
\hline
\end{tabular}
\end{table}
The five statements were repeated in the post-study questionnaire to assess changes in these measures for RQ1.

Participants then experienced each of the three control methods. To reduce systematic ordering effects, the presentation order was counterbalanced using the six possible orderings of \textit{Joystick}, \textit{Waypoint}, and \textit{Path-Drawing}. Participants were distributed across these orders as evenly as possible.

For each control method, participants completed a one-minute interface familiarisation period followed by approximately one minute of practice controlling the physical robot with each control method without the virtual animal task. They then completed a two-minute interaction for each control method within the animal-based MR activity.

After completing all three control methods, participants completed a post-study questionnaire assessing perceived workload, engagement, and control-method preference. 

Perceived workload was assessed using an adapted comparative-ranking measure based on the six dimensions of the NASA Task Load Index (NASA-TLX): mental demand, physical demand, temporal demand, performance, effort, and frustration~\cite{sandra_g_hart_development_1988}.

Engagement was assessed using four statements adapted from the dimensions of the User Engagement Scale Short Form (UES-SF): focused attention, perceived usability, aesthetic appeal, and reward~\cite{OBrien2018UES}. One statement was used for each dimension to limit questionnaire length and repetition within the short study, as shown in Table~\ref{tab:engagement_statements}.
\begin{table}[ht]
\centering
\caption{Statements used to assess engagement across control methods.}
\label{tab:engagement_statements}
\small
\begin{tabular}{p{0.4cm} p{1.4cm} p{5.5cm}}
\hline
\textbf{ID} & \textbf{Category} & \textbf{Statement} \\
\hline
E1 & Focused Attention & I became deeply absorbed in the experience while using this control method and lost track of the time passing. \\
E2 & Perceived Usability & I found this control method easy to understand and use without feeling confused, frustrated, or overly taxed. \\
E3 & Aesthetic Appeal & I found the overall look, feel, and presentation of this control method appealing. \\
E4 & Reward & I found using this control method was interesting, worthwhile, and rewarding. \\
\hline
\end{tabular}
\end{table}

For each of the ten perceived workload and engagement measures, participants ranked \textit{Joystick}, \textit{Waypoint}, and \textit{Path-Drawing} controls relative to each other. Participants also selected their preferred control method and answered open-ended questions about the reasons for their preferences and aspects of the application that influenced their experience. Responses to the open-ended questions were recorded and transcribed for qualitative analysis.

\subsection{Analysis Methodology}
\label{sec:analysis}

\subsubsection{RQ1: Change in Interest and Confidence}

\begin{figure*}[t]
    \centering
    \begin{tabular}{cc}
    \includegraphics[width=0.45\textwidth,trim={0cm 6.5cm 50cm 0cm}, clip]{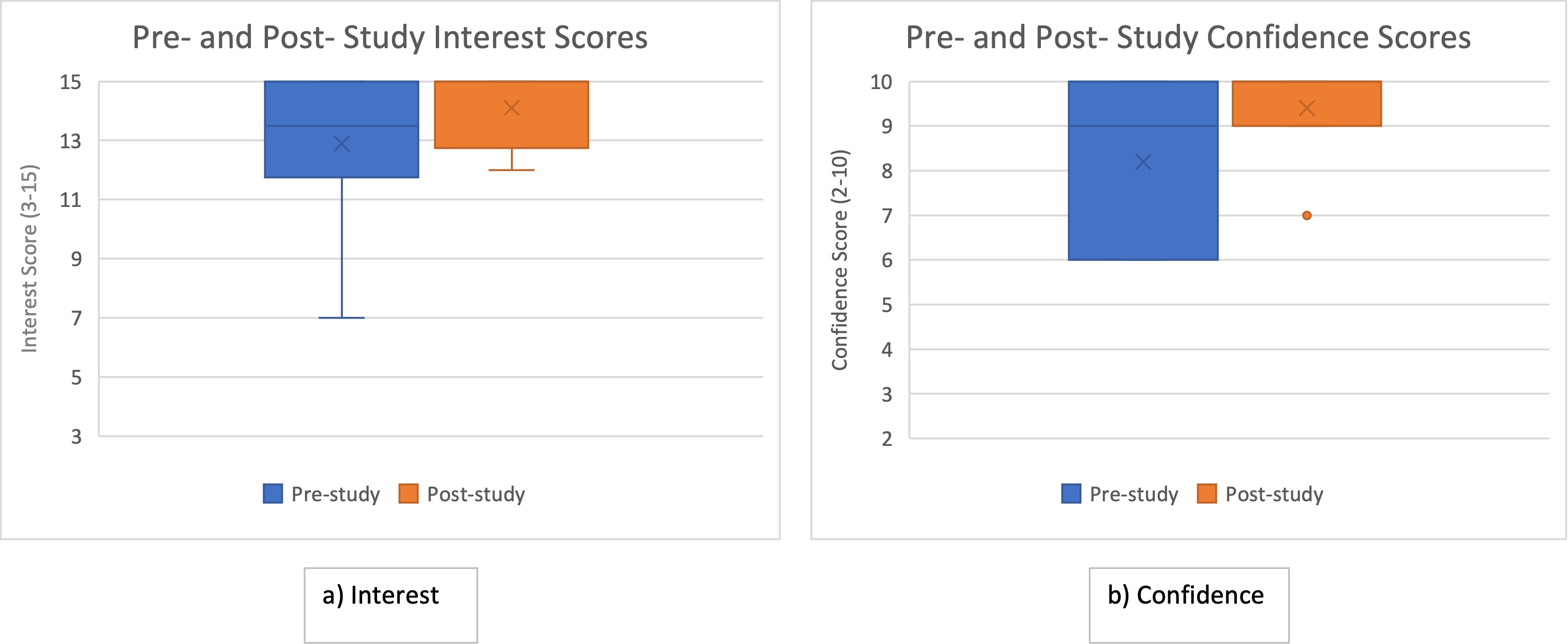}&
    \includegraphics[width=0.45\textwidth,trim={50cm 6.5cm 0cm 0cm}, clip]{figures/distribution_pre_post.jpg}\\
    (a) & (b) \\
    \end{tabular}
    \caption{Distribution of pre- and post-study scores for (a) interest and (b) confidence.}
    \label{fig:pre_post_scores}
\end{figure*}

The three interest items were combined to produce an overall interest score, and the two confidence items were combined to produce an overall confidence score as shown in Figure~\ref{fig:pre_post_scores}. Given the ordinal questionnaire responses and small sample size, pre- and post-study scores were compared separately using Wilcoxon signed-rank tests. Interest and confidence were treated as distinct, pre-specified outcomes, and the reported $p$-values were not adjusted for multiple comparisons. These analyses were therefore interpreted as exploratory.

Differences between participants with and without prior robotics experience were examined descriptively due to the small subgroup sizes ($n=5$).

\label{sec:rq1_analysis}

\subsubsection{RQ2: Control-Method Comparison}

Friedman tests were used to determine whether participants' rankings differed between \textit{Joystick}, \textit{Waypoint}, and \textit{Path-Drawing} control for each engagement and workload measure, with a significance threshold of $\alpha=0.05$. Holm correction was applied separately across the six workload tests and across the four engagement tests. Kendall's $W$ was calculated as a measure of effect size. Where a significant overall difference was identified, pairwise sign tests were used for post-hoc comparisons, with Holm correction applied across the three comparisons for that measure. Rank coding followed the direction of each question: higher ranks represented more of the named construct, except for Performance, where higher ranks represented worse perceived performance. Control-method preference was summarised using selection frequencies.

Differences between participants with and without prior robotics experience were examined descriptively due to the small subgroup sizes ($n=5$).

\subsubsection{Qualitative Analysis}

Open-ended responses were analysed using inductive thematic analysis. Relevant excerpts were coded using a codebook with definitions, inclusion and exclusion criteria, and representative examples, then grouped into themes. These themes contextualised the quantitative findings.

\section{Results}

\subsection{RQ1: Change in Interest and Confidence}

Pre- and post-study scores were compared to determine whether participants' interest and confidence changed following the MR robotics experience. The combined interest score increased from $12.9 \pm 2.64$ before the experience to $14.1 \pm 1.29$ after the experience (mean $\pm$ SD); however, this change was not statistically significant ($p=0.125$). The median increased from 13.5 to 15 (out of 15). The combined confidence score increased from $8.2 \pm 1.75$ to $9.4 \pm 0.97$, with the Wilcoxon test indicating a significant increase using the unadjusted significance threshold ($p=0.031$). The median increased from 9 to 10 (out of 10).

All five participants without prior robotics experience showed a change in at least one interest or confidence item, compared with one of five participants with prior robotics experience. No participant reported a decrease in any of the five statements. However, participants with prior experience had high initial scores, with median pre-study scores of 15 for interest and 10 for confidence.

\subsection{RQ2: Control Method Engagement and Workload}
\label{sec:rq2_results}

\begin{table}[t]
\centering
\caption{Post-hoc pairwise comparisons for dimensions with significant Friedman tests (J: Joystick; W: Waypoint; P: Path-Drawing). Bold values indicate significance following Holm correction.}
\label{tab:posthoc}
\small
\setlength{\tabcolsep}{1.5pt}
\renewcommand{\arraystretch}{1.05}
\begin{tabular}{lccc}
\hline
\textbf{Measure} & \shortstack{\textbf{Mean Rank}\\\textbf{(J/W/P)}} & \textbf{Comparison} & $\mathbf{p_{\mathrm{adj}}}$ \\
\hline
Mental Demand & 2.2/1.3/2.8 & J--W & 0.219 \\
 & & J--P & 0.344 \\
 & & W--P & \textbf{0.006} \\[2pt]
Physical Demand & 2.0/1.5/2.7 & J--W & 0.688 \\
 & & J--P & \textbf{0.043} \\
 & & W--P & \textbf{0.006} \\[2pt]
Temporal Demand & 2.1/1.6/2.3 & J--W & 0.688 \\
 & & J--P & 0.688 \\
 & & W--P & 0.344 \\[2pt]
Performance & 1.1/2.0/2.9 & J--W & \textbf{0.043} \\
 & & J--P & \textbf{0.006} \\
 & & W--P & \textbf{0.043} \\[2pt]
Effort & 2.2/1.4/2.8 & J--W & 0.219 \\
 & & J--P & 0.344 \\
 & & W--P & \textbf{0.006} \\[2pt]
Frustration & 1.1/1.9/3.0 & J--W & \textbf{0.043} \\
 & & J--P & \textbf{0.006} \\
 & & W--P & \textbf{0.043} \\[2pt]
Perceived Usability & 2.3/2.7/1.0 & J--W & 0.344 \\
 & & J--P & \textbf{0.006} \\
 & & W--P & \textbf{0.006} \\
\hline
\end{tabular}
\end{table}

%\begin{figure*}[!t]
%    \centering
%        \begin{tabular}{cc}
%    \includegraphics[width=0.4\textwidth,trim={0cm 4cm 42cm 0cm}, clip]{figures/tlx_rankings.jpg}&
%    \includegraphics[width=0.4\textwidth,trim={42cm 4cm 0cm 0cm}, clip]{figures/tlx_rankings.jpg}\\
%    (a) & (b) \\
%    \end{tabular}
%    \caption{Distribution of participant rankings across control methods for workload dimensions adapted from NASA-TLX: (a) Performance and (b) Frustration.}
%   \label{fig:tlx_rankings}
%\end{figure*}

\begin{figure*}[t]
    \centering
    \includegraphics[width=\textwidth]{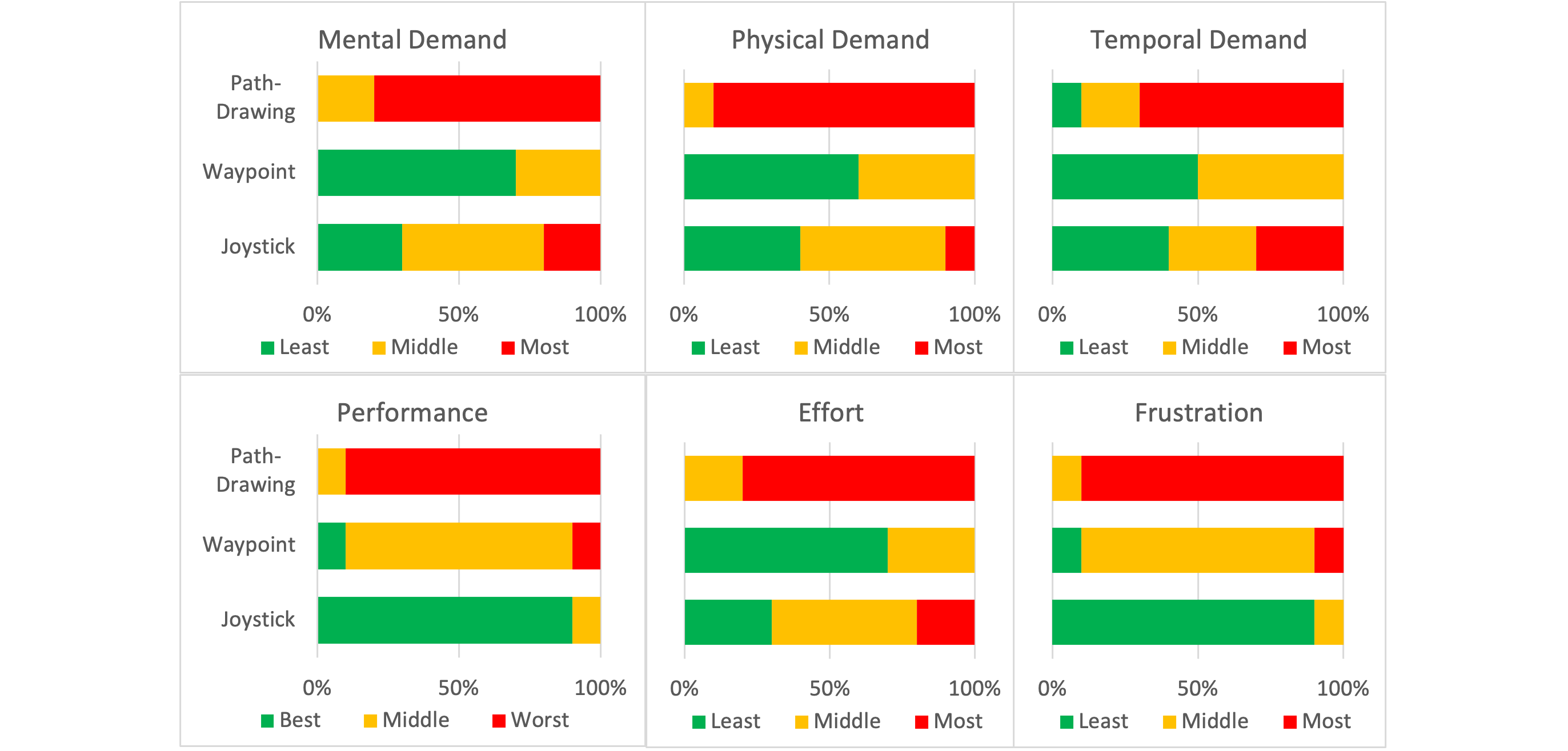}
    \caption{Distribution of participant rankings across control methods for the six workload dimensions adapted from NASA-TLX.}
    \label{fig:tlx_rankings}
\end{figure*}

\subsubsection{Perceived Workload}

Significant differences between the three control methods were identified across all six workload dimensions following Holm correction: Mental Demand ($Q=11.4$, $p_{\mathrm{adj}}=0.010$, $W=0.57$), Physical Demand ($Q=12.6$, $p_{\mathrm{adj}}=0.007$, $W=0.63$), Temporal Demand ($Q=6.2$, $p_{\mathrm{adj}}=0.045$, $W=0.31$), Performance ($Q=16.2$, $p_{\mathrm{adj}}=0.002$, $W=0.81$), Effort ($Q=11.4$, $p_{\mathrm{adj}}=0.010$, $W=0.57$), and Frustration ($Q=16.2$, $p_{\mathrm{adj}}=0.002$, $W=0.81$).

\textit{Path-Drawing} was generally associated with greater workload, being ranked as the most mentally demanding by nine participants, most physically demanding by seven, most effortful by eight, and most frustrating by nine. Post-hoc comparisons showed that \textit{Path-Drawing} was significantly more mentally demanding and effortful than \textit{Waypoint}, and more physically demanding than both \textit{Waypoint} and \textit{Joystick} (Table~\ref{tab:posthoc}). Although Temporal Demand differed overall, no pairwise comparison remained significant after correction. The largest effects were observed for Performance and Frustration ($W=0.81$ for both). Nine participants ranked \textit{Joystick} as producing the best performance and \textit{Path-Drawing} as producing the worst, with significant differences between all three methods. Similarly, nine participants ranked \textit{Joystick} as the least frustrating and \textit{Path-Drawing} as the most frustrating, again with significant differences between all three methods (Table~\ref{tab:posthoc}, Figure~\ref{fig:tlx_rankings}).

\begin{figure*}[!t]
    \centering
    \includegraphics[width=\textwidth]{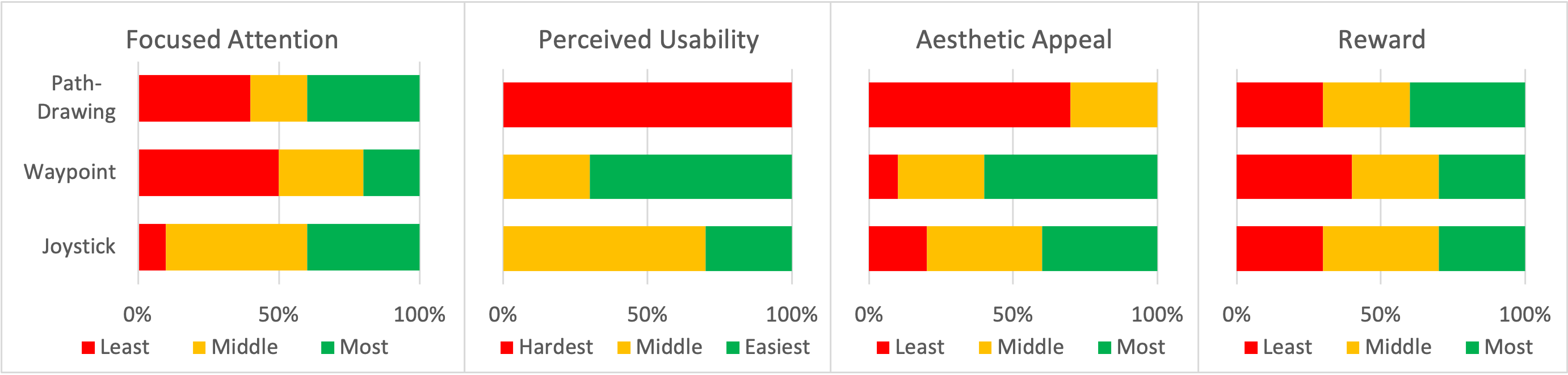}
    \caption{Distribution of participant rankings across control methods for the four engagement dimensions adapted from UES-SF.}
    \label{fig:ues_rankings}
\end{figure*}

%\begin{figure*}[!t]
%    \centering
%    \begin{tabular}{cc}
%    \includegraphics[width=0.4\textwidth,trim= {0cm 3.45cm 42cm 0cm}, clip]{figures/ues_rankings.jpg}&
%    \includegraphics[width=0.4\textwidth,trim={42cm 3.45cm 0cm 0cm}, clip]{figures/ues_rankings.jpg}\\
%    (a) & (b) \\
%    \end{tabular}
%    \caption{Distribution of participant rankings across control methods for engagement dimensions adapted from UES-SF: (a) Perceived Usability and (b) Reward.}
%    \label{fig:ues_rankings}
%\end{figure*}

\subsubsection{Engagement and Preference}

Of the four engagement dimensions, only Perceived Usability differed significantly between control methods following Holm correction ($Q=15.8$, $p_{\mathrm{adj}}=0.001$, $W=0.79$). No significant differences were identified for Focused Attention ($Q=1.8$, $p_{\mathrm{adj}}=0.813$, $W=0.09$), Aesthetic Appeal ($Q=7.8$, $p_{\mathrm{adj}}=0.061$, $W=0.39$), or Reward ($Q=0.2$, $p_{\mathrm{adj}}=0.905$, $W=0.01$).

All participants ranked \textit{Path-Drawing} as the hardest method to use, while seven ranked \textit{Waypoint} as the easiest. Post-hoc comparisons showed that both \textit{Joystick} and \textit{Waypoint} were significantly easier to use than \textit{Path-Drawing}, with no significant difference between \textit{Joystick} and \textit{Waypoint} (Table~\ref{tab:posthoc}, Figure~\ref{fig:ues_rankings}).

Despite no significant differences in Focused Attention, Aesthetic Appeal, or Reward, eight participants (80\%) selected \textit{Joystick} as the preferred control method, while the remaining two selected \textit{Waypoint} (20\%). No participant selected \textit{Path-Drawing}.

\subsubsection{Prior Robotics Experience}

Overall preference was identical between prior-experience groups: four of the five participants in each group preferred \textit{Joystick} and one preferred \textit{Waypoint}. Descriptively, both groups also showed similar workload patterns, with \textit{Path-Drawing} generally receiving the highest workload rankings.

\subsection{Qualitative Findings}
\label{sec:qualitative_results}

The thematic analysis identified six recurring themes: \textit{Direct Control}, \textit{Simple Autonomous Control}, \textit{Path-Drawing Challenges}, \textit{Purposeful Interaction}, \textit{Physical--Virtual Connection}, and \textit{Technical Barriers}.

\textit{Direct Control} reflected experiences with \textit{Joystick}, which was commonly described as easy, predictable, and familiar, with some participants relating this familiarity to previous gaming experience. \textit{Simple Autonomous Control} reflected experiences with \textit{Waypoint}, which was similarly viewed as easy to use and required less continuous input. However, this reduced involvement was less engaging for some participants, with one describing it as ``a bit boring'' because it was ``just point and click.'' \textit{Path-Drawing Challenges} captured difficulties with the drawing interaction and robot behaviour, including hand coordination, the robot becoming stuck, and movement differing from the intended path. Despite these difficulties, some participants found the concept novel and interesting.

\textit{Purposeful Interaction} reflected how the animal-based task provided an objective beyond simply navigating the robot, with participants describing the animals as giving the activity purpose and becoming emotionally invested in the task. \textit{Physical--Virtual Connection} captured positive responses to interacting with virtual elements while controlling a physically present robot. Finally, \textit{Technical Barriers} included visual performance, headset discomfort, and difficulty operating the robot. Some participants also reported becoming more comfortable with the headset and interactions as the activity progressed.

\section{Discussion}
\label{sec:discussion}

\subsection{Interest and Confidence in Robotics}

Confidence increased following the MR robotics experience, with the unadjusted Wilcoxon test reaching statistical significance ($p=0.031$). This result is consistent with the confidence component of the RQ1 hypothesis and provides preliminary evidence that an MR robotics experience may support participants' confidence in engaging with robotics. 

Interest also increased following the experience, but the difference was not statistically significant ($p=0.125$). High initial interest scores may have created a ceiling effect, with several participants beginning at or near the maximum possible score and therefore having limited opportunity to demonstrate an increase. This may have reduced the measure's sensitivity to changes in interest and provides important context for the result.

More participants without prior robotics experience showed changes in at least one interest or confidence item than those with prior experience, whose high baseline scores left less scope for improvement. While consistent with the hypothesis, the small subgroup sizes ($n=5$) mean this pattern should be interpreted cautiously.

\subsection{Control Method Experience}

Workload differed significantly across all six adapted NASA-TLX dimensions, with the largest differences in Performance and Frustration ($W=0.81$ for both). Overall, \textit{Joystick} was associated with better perceived performance and lower frustration, while \textit{Path-Drawing} was generally the most demanding. The \textit{Direct Control}, \textit{Simple Autonomous Control}, and \textit{Path-Drawing Challenges} themes suggest that these differences may reflect how reliably participants translated their intentions into robot behaviour. In particular, implementation difficulties with \textit{Path-Drawing} may have contributed to its higher workload.

Engagement differed less between the methods, with Perceived Usability the only dimension showing a significant difference ($W=0.79$). Notably, Reward rankings were evenly distributed across the three methods (see Figure~\ref{fig:ues_rankings}), even though eight of ten participants selected \textit{Joystick} as their preferred method. This suggests that preference was not determined by how interesting or rewarding a method was alone. Qualitative responses provide further context, with participants identifying different engaging qualities across the methods, including the novelty of \textit{Path-Drawing}, despite its usability difficulties. Overall, the results of RQ2 indicate that the control methods differed more clearly in workload, usability, and preference than in broader engagement.

Furthermore, prior robotics experience did not correspond with the hypothesised difference in preference, with the same preference pattern observed in both groups; however, conclusions are limited given the small subgroup sizes.

\subsection{Implications for Designing MR Robotics Experiences}

Taken together, the workload, engagement, and preference findings suggest that control methods can contribute different qualities to an introductory robotics experience. A method that is easy to use or preferred is not necessarily more engaging across all dimensions, while a more demanding interaction may still offer novelty or interest. Rather than selecting a control method based solely on preference, the intended experience should therefore be considered when choosing how participants interact with the robot. Providing multiple forms of control may also expose participants to different aspects of robotics, provided that usability limitations do not interfere with the intended interaction.

The themes relating to \textit{Purposeful Interaction} and \textit{Physical--Virtual Connection} further suggest that the control interface was only one component of the experience. The animal-based objective gave robot movement purpose, while integrating virtual content with a physically present robot connected the MR activity to observable robot behaviour. These findings suggest that introductory MR robotics activities may benefit from situating robot control within a meaningful task rather than presenting control as an isolated technical exercise.

Given the project's motivation to engage girls with robotics, these findings should not be interpreted as identifying a control method that girls inherently prefer. Instead, they suggest the value of creating accessible experiences that let participants interact directly with robotics without requiring programming knowledge, while providing purposeful tasks and different forms of interaction. The \textit{Technical Barriers} and \textit{Path-Drawing Challenges} themes also highlight the importance of interaction reliability, as technical difficulties may shift participants' attention away from the intended robotics experience. 

\section{Limitations and Future Work}
\label{sec:limitations}

The study was limited by its small sample of ten participants, including only five participants in each prior-experience group, meaning subgroup patterns should be interpreted cautiously. The exploratory sample was not designed or powered to assess gender differences, and the findings should not be interpreted as specific to girls or women. High baseline interest and confidence also produced a ceiling effect that may have limited measurable change for RQ1. The control methods differed in input modality, autonomy, implementation reliability, and control abstraction; therefore, differences between methods cannot be attributed to control abstraction alone. Future work should evaluate the experience with a larger and more varied sample, particularly girls with limited prior robotics experience, and consider longer-term measures of changes in robotics interest and confidence.

Technical difficulties with \textit{Path-Drawing}, particularly hand tracking and path execution, also make it difficult to separate responses to the control concept from its implementation. Future work should therefore re-evaluate \textit{Path-Drawing} after improving its reliability. Longer interactions and additional robotics tasks could also be investigated to determine whether the findings extend beyond the short, single-task experience evaluated in this study.

\section{Conclusion}
\label{sec:conclusion}

This work demonstrates how MR can combine virtual task elements with physical robot interaction to create an accessible introductory robotics experience. The observed increase in participant confidence provides preliminary evidence that such experiences could support people with limited robotics exposure, although further evaluation with a larger sample is required. The comparison of control methods showed differences in workload, usability, and preference, whereas no statistically significant differences were detected in the other engagement dimensions. While \textit{Joystick} was strongly preferred, participants found engaging qualities across the different methods, demonstrating that preference alone does not capture the complete interaction experience. These findings highlight the value of designing introductory MR robotics activities around reliable interaction, meaningful tasks, and control methods selected according to the experience they are intended to provide.

\section*{Acknowledgement}

The authors thank UTS:RI for providing facilities and acknowledge support from the ARC Industrial Transformation Training Centre for Collaborative Robotics in Advanced Manufacturing (IC200100001).

OpenAI Codex was used to organise author-provided materials into an early scaffold of the first few sections and later for language polishing. The authors substantially rewrote and verified the entire manuscript. No AI system generated or modified the study data or statistical results.

\bibliography{library}
\bibliographystyle{named}

\end{document}